# Catalyst-free MBE growth of PbSnTe nanowires with tunable aspect ratio

Mathijs G.C. Mientjes[1,*], Xin Guan[1,*,⊥], Pim J.H. Lueb[1], Marcel A. Verheijen[1,2],

Erik P.A.M. Bakkers[1, ⊥]

[1]*Eindhoven University of Technology, 5600MB Eindhoven, The Netherlands*

[2]*Eurofins Materials Science Eindhoven, 5656AE Eindhoven, The Netherlands*

[*] These authors contributed equally to this work.

[⊥] Corresponding authors: guanxin246@gmail.com, ebakkers@tue.nl

**Abstract**

Topological crystalline insulators (TCIs) are interesting for their topological surface states, which hold great promise for scattering-free transport channels and fault-tolerant quantum computing. A promising TCI is SnTe. However, Sn-vacancies form in SnTe, causing a high hole density, hindering topological transport from the surface being measured. This issue could be relieved by using nanowires with a high surface-to-volume ratio. Furthermore, SnTe can be alloyed with Pb reducing the Sn-vacancies while maintaining its topological phase. Here we present the catalyst-free growth of monocrystalline PbSnTe in molecular beam epitaxy (MBE). By the addition of a pre-deposition stage before the growth, we have control over the nucleation phase and thereby increase the nanowire yield. This facilitates tuning the nanowire aspect ratio by a factor of four by varying the growth parameters. These results allow us to grow specific morphologies for future transport experiments to probe the topological surface states in a $Pb_{1-x}Sn_xTe$-based platform.

At the boundary between a topological insulator (TI) and a trivial material, conducting surface states appear [1], surrounding the insulating bulk. These surface states are of interest due to their potential for dissipation-less and spin-polarized transport [1] [2] [3]. Additionally, these states are topologically protected, making them robust against disorder [2]. TCIs are a class of TIs, where the topology arises from the crystal mirror symmetries [4]. Surface states on a TCI are therefore well-defined and expected to be only sensitive to the breaking of this symmetry by, for example, lattice strain or ferroelectric coupling [5]. SnTe has been predicted theoretically to be a TCI [6] and its number of Dirac cones varies depending on the crystal plane of the facet. Signatures of surface states on SnTe {100} and {111} facets have been shown experimentally using angle-resolved photo-emission spectroscopy (ARPES) [7] [8]. To further investigate the topological properties of the surface states, transport measurements are required. It has, however, been proven challenging to probe these surface states in transport due to the high hole concentration in the bulk dominating the transport properties [9] [10] [11] [12]. This issue could be mitigated by choosing a geometry with a high surface-to-volume ratio, such as nanowires (NWs) where the surface states are expected to be more pronounced in transport experiments [13]. Additionally, the carrier density can be reduced by alloying SnTe with Pb, while maintaining the topological phase for x > 0.4 in $Pb_{1-x}Sn_xTe$ [10] [14]. This calls for high-quality PbSnTe NWs with tunable composition and well-defined facets. Au-catalyzed (Pb)SnTe NWs have been grown in both chemical vapor deposition and MBE [15] [16] [17]. However, control over the NW morphology has proven difficult. Furthermore, incorporation of foreign atoms from the catalyst particle into the NW might obfuscate the electronic transport results as has been observed for other material systems [18]. Here, we develop the growth of catalyst-free PbSnTe NWs which are formed by selective-area growth in a vapor-solid fashion by MBE. The NWs are defined by atomically flat {100} side and top facets. The wires grow from patterned holes in a mask. By varying the hole size (HS) and the distance between holes (pitch) we show control of the NW morphology, and most importantly the aspect ratio, while maintaining a high crystal quality.

The $Pb_{1-x}Sn_xTe$ NWs are grown on [100]-oriented Si substrates with a 20 nm $SiN_x$ mask. The holes in the mask are patterned within fields using electron-beam lithography and etched using reactive ion etching. Growth occurs in an ultra-high vacuum (background pressure of 3-$6 \times 10^{-10}$ mBar) MBE system. The substrate temperature is monitored using the band edge absorption (BandiT) of a GaAs reference substrate. Material is deposited from Pb and Sn effusion cells and a Te cracker cell. The fluxes are measured in beam equivalent pressure (BEP) by a beam flux monitor before the growth. The total flux of material at the substrate is denoted as $F_{tot}$. All presented samples are grown with a group IV to group VI ratio of the fluxes of 1/3 and a Pb to Sn ratio of 1 ($x_{input}$=0.5). All material fluxes have a 30-degree incident angle to the substrate's normal. The substrates are kept rotating at 5.3 rpm. Unless otherwise mentioned, the growth time is 220 minutes. Further growth details can be found in the supplementary information (SI 1).

**Results**

We start exploring the basic growth as a function of total flux and substrate temperature as shown in Figure 1(a). A high yield (> 50%) of NWs can be obtained in a small growth regime, indicated by the white area in Figure 1(a). Here, we define yield as the fraction of patterned holes containing nanowires, as described in SI 2. Outside of this growth regime (Figure 1(a) grey area), either parasitic growth dominates at lower temperatures and/or higher fluxes, or there is no nucleation at all at higher temperatures and/or lower fluxes. This is due to a critical balance between the incoming flux and evaporation of adatoms from the substrate before they contribute to crystal growth [19]. All further growth runs presented in this work are performed at T=339 °C and $F_{tot}$ =2x10$^{-7}$ BEP (circled data point). It should be noted that this temperature is significantly lower than previously reported for the growth of (Pb)SnTe nanostructures [15] [16] [17]. This could prove beneficial for limiting the formation of Sn-vacancies [9].

As can be seen in the scanning electron microscopy (SEM) image of Figure 1(b), two types of structures can be distinguished, NWs and nanoflakes. In this work, we will focus on the growth of NWs, and discuss the nanoflakes elsewhere. NWs grow approximately vertically along a <100> direction with an average tilt angle of less than 10 degrees (SI 2). NWs have full rotational freedom around the vertical <100> axis indicating that there is no epitaxial relationship between the wires and the substrate. We attribute this to the presence of a native oxide on the silicon substrate in the patterned holes before NW growth. Figure 1(c) shows a typical bright field transmission electron microscopy (BFTEM) image of a nanowire with a zoom-in in Figure 1(d). The NWs have atomically flat {100} side and top facets. Cross-sectional High Angle Annular Dark Field scanning TEM (HAADF-STEM) images confirm the rock salt crystal structure with no observable defects (Figure 1 (e,f)). In addition, detailed side-view inspection along the entire length of many NWs did not reveal the presence of any structural defects. A 2-3 nm self-limiting oxide layer is formed on the surface, which predominantly consists of $SnO_x$ as determined by energy-dispersive X-ray spectroscopy (EDX) (SI 3). The corners are slightly rounded, probably due to a minimization of the surface energy. We note that the growth is reproducible regarding the NW yield/dimensions and flake formation.

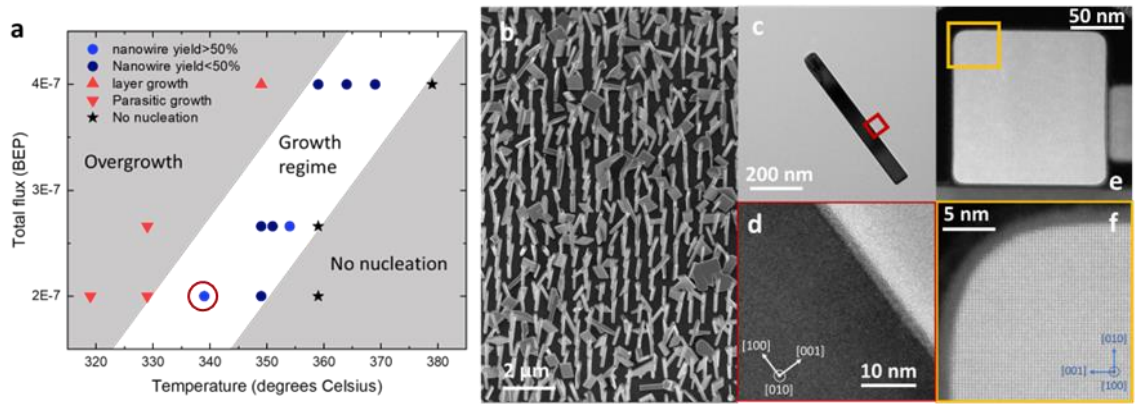

Fig 1: (a) The growth parameter space depicted by plotting the average nanowire yield of the PbSnTe NWs grown in a HS of 30 nm and a pitch of 500 nm versus the growth temperature and the total flux. (b) A representative SEM image of selectively grown PbSnTe NWs on a [100] Si substrate with a SiN$_x$ mask. (c) low-resolution BFTEM image of a PbSnTe nanowire (d) high-resolution TEM image of the side facet imaged along the [010] zone axis. All surface facets are of the {100} family. A 2-3 nm native oxide layer is present on all facets. (e) cross-sectional HAADF STEM image showing the square cross-section. (f) high-resolution HAADF-STEM image showing the cubic crystal structure without observable defects.

The NW growth data presented above has been obtained using an optimized nucleation procedure. An essential part of the procedure is a pre-deposition step of the group IV material before NW growth to increase the NW yield. Four types of pre-deposition are studied: i) No pre-deposition, ii) 2 minutes of simultaneous Sn and Pb pre-deposition, iii) 4 minutes of Sn and iv) 4 minutes of Pb. To distinguish between the role of Sn and Pb, we studied both elements individually in iii) and iv). The fluxes of Sn and Pb correspond to 2.5x10$^{-8}$ BEP in all experiments. To account for the halved total flux in cases III and IV, the pre-deposition time of these experiments was doubled. To separately study the effects of nucleation and growth, first, the result of the pre-deposition runs is investigated and SEM images are shown in the top panels of Figure 2. Next, these pre-depositions are used to grow NWs and resulting SEM images are shown in the bottom panels of Figure 2.

Without a pre-deposition step before the growth (i), no droplets can be seen within and around the patterned holes (Figure 2a). The nucleation yield, which we define as the fraction of holes with a faceted object, after growth is >99% and the nanowire yield is $27 \pm 2\%$, with a nanowire-to-flake ratio of $0.5\pm0.2$. After the Sn and Pb pre-deposition step (ii), small droplets can be found within the patterned hole and on the $SiN_x$ mask. From XPS analysis it is found that these droplets consist mainly of Sn (See SI 5), which is expected from the difference in vapor-pressure of Sn and Pb [19]. The nucleation yield is >99%, and the NW yield has increased to $52 \pm 1\%$ (Fig. 2b)), resulting in a nanowire-to-flake ratio of $1.8\pm0.4$. For the 4 minutes of Sn pre-deposition (iii), the size of the droplets within and around the patterned holes has increased (Fig. 2c). This agrees with the observations in type (ii) that Sn has a low vapor pressure and will stick to the mask and substrate while the Pb tends to evaporate. A double dose of Sn (case ii vs iii) leads to more accumulation of Sn on the substrates. Again, a nucleation yield of >99% is observed, The NW yield and nanowire-to-flake ratio are $41 \pm 1\%$ and $0.9\pm0.2$ respectively. For the 4 minutes of Pb deposition (iv), no droplets are observed within and around the patterned holes (Fig. 2d). The nucleation yield and NW yield have decreased drastically to <5% and $0.2\pm0.1\%$, respectively. This indicates that the Pb alters the surface properties of the substrate and enhances evaporation from the mask thereby decreasing the nucleation yield.

The pre-deposition does not enhance the nucleation yield for cases i-iii but instead promotes nanowire over nanoflake formation. The relation between the pre-deposition and the nanowire-to-flake ratio is not yet fully understood, but we hypothesize that it is related to the position where the nucleation occurs within the patterned hole. If no droplets are present, nucleation might tend to start at the edge of a hole. This asymmetry might promote flake over nanowire formation. In the case of a short pre-deposition step, small droplets are formed within the hole. In this case, the nucleation geometry is symmetric and nanowire growth is promoted. If more material is added to the hole (i.e. case iii), Sn droplets may start to contact the edge of the mask and now the nucleation geometry is again asymmetric, promoting nanoflake formation. A final remark on the nucleation connects to the observation that no parasitic growth on the mask is observed in the lower panels of Figure 2, even though droplets are discernable on the mask surface in the upper panels of Figure 2. This implies that the nucleation probability of PbSnTe crystals is higher on the $SiO_x$ layer in the holes than on the $SiN_x$ mask. The Sn/Pb droplets formed on the mask during the pre-deposition step disappear during the later stages of growth. These may have evaporated or have released adatoms that diffuse on the mask contributing to NW growth.

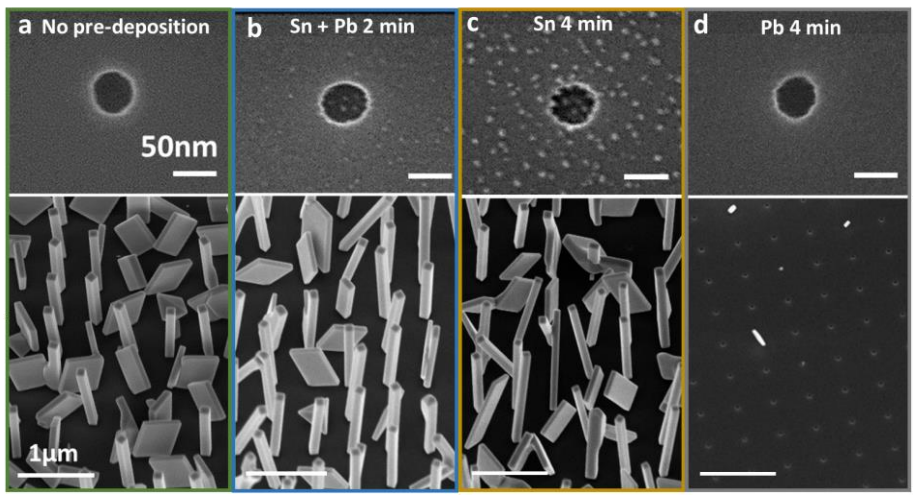

Fig 2: Representative SEM images of the PbSnTe growth substrate before (top panels) and after growth of PbSnTe nanostructures (bottom panels) grown at a pitch of 500 nm and a hole size of 30 nm. (Top panels) The mask openings after the following pre-deposition steps a) no pre-deposition, b) 2 minutes of Pb and Sn pre-deposition, c) 4 minutes of Sn pre-deposition and d) 4 minutes of Pb pre-deposition, respectively. (Bottom panels) the samples after growth with the respective pre-depositions.

Now that we have achieved reproducible growth with a relatively high yield, we would like to reveal the role of the pre-deposited material and understand the growth mechanism in the early stages. Therefore, structures have been grown for short times and the results are shown in Figure 3 for a hole size (HS) of 20 nm and 100 nm. After 2 minutes of Sn and Pb pre-deposition and 15 minutes of subsequent growth, a small seed crystal is observed within the holes with a 20 nm diameter (Fig. 3a). For the larger holes, at 15 minutes multiple seed crystals are visible with varying shapes and orientations (Fig. 3c). This indicates that the Sn/Pb droplets act as collection spots for the incoming Te which has a high vapor pressure at this temperature and would quickly evaporate from the mask when unbound [19]. Upon absorption of Te these metallic Sn/Pb droplets crystallize into small PbSnTe crystals. In larger holes, more droplets are present resulting in more seed crystals. After 60 minutes, faceted objects can be observed that reflect the morphology of nanowires and flakes (Fig. 3b). From their faceting, it can be concluded that these objects are single-crystalline, implying that between 15 and 60 minutes of growth the merging of the various seed crystals is accompanied by recrystallization and/or reorientation of most of the seed crystals [20].

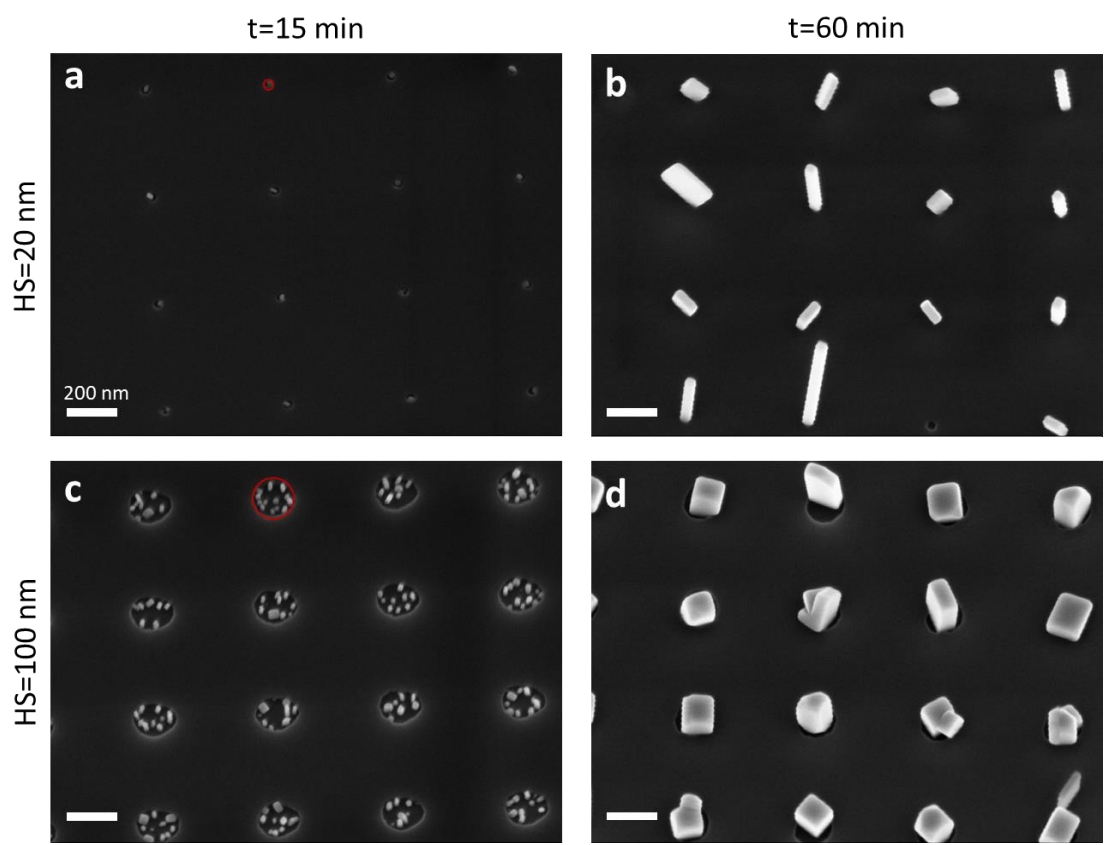

Fig 3: Representative SEM images of the substrate after 15 minutes (a,c) and 60 minutes (b,d) of growth with a hole size (HS) of 20 nm (a,b) and 100 nm (c,d). All images were taken at identical magnification. In Fig 3 (a,c) the hole size is indicated with a red circle.

To understand if the NWs continue to grow via a Vapor-Liquid-Solid (VLS) mechanism, in which the group IV acts as the catalyst, or rather via a Vapor-Solid (VS) mechanism, we have carried out additional growth experiments. The results of these experiments are schematically depicted in SI6 and can be summarized as follows: Firstly, a droplet has never been observed on top of the NWs in SEM or TEM after growth. Secondly, if a 2 min Sn/Pb deposition is performed after NW growth, there are droplets visible on the NW facets showing that the droplets do not evaporate during the cooldown. Then, if growth is resumed after this Sn/Pb deposition, the growth continues as if there has been neither interruption nor metal deposition. Most importantly, in the latter experiment NW growth has not been nucleated from the Sn/Pb deposits on the initial NW side facets, which would lead to the formation of branches in case of VLS growth [21]. As a final experiment, the growth is interrupted by a 10 min Te deposition and then resumed, resulting in continued NW growth as without interruption. If the NWs would grow via a VLS mechanism, the Sn(Pb) droplet would be consumed during Te exposure, terminating further NW growth. Based on these observations, we conclude that after 60 minutes, the growth is dominated by the VS mechanism.

With a VS growth mode, with solely surfaces from the same crystal family, it would be expected, in first instance, that the radial and axial growth rates are identical, resulting in cube-shaped crystals. This is, however, not observed as already shown in Figure 1(b). To explain this apparent paradox, the NW growth kinetics are studied in more detail by varying the pitch (in the range 250-2000 nm) using arrays with a HS of 30 nm. The average length, width and the resulting aspect ratio of these NWs are presented in Figure 4 (a,b). The NW width increases with increasing pitch from 64 nm to 213 nm. In this range, the NW length remains relatively constant, resulting in a decrease in the aspect ratio by a factor of 4. The aspect ratio seems to approach a constant value for the largest pitches. In addition, the effect of HS on the final NW dimensions is studied using a pitch of 500 nm, and a nominal HS ranging from 20 to 100 nm. From Figure 4 (c,d), it can be seen that the NW width increases with HS while the length decreases with increasing HS, resulting in a decrease of the aspect ratio by a factor of 2.7.

The pitch-dependence of the NW width can be understood by considering shadowing from nearby nanowires as well as diffusion of adatoms on the $SiN_x$ mask [22] [23] [24]. Material fluxes in MBE are directional, and NWs will be shadowed by surrounding NWs for the smallest pitches, limiting the amount of material deposited on the side facets. In addition, adatoms can arrive on the mask and diffuse to nearby NWs. For smaller pitches, the collection areas of the NWs overlap leading to competition [23], resulting in a smaller amount of material reaching the NW side facets, and a lower radial growth rate. The fact that the axial growth rate is not affected by varying the pitch indicates that material supply from the substrate mask and the NW side facets to the top of the NW is limited. In other words, axial growth is mainly due to direct impingement. The non-unity aspect ratio of the nanowires, obtained for large pitches, where shadowing and competition effects are not relevant, can be explained by the importance of direct impingement in combination with the (30°) angle of the incoming fluxes given the MBE chamber configuration. This leads to more material supply to the top facet than to the side facets. Furthermore, due to the rotation of the substrate, the side facets are only exposed to the source flux for a fraction of the time while the top facet is constantly exposed. For infinite growth times, the aspect ratio for every HS is expected to converge to a value of 2.7 (SI 7). However, for a growth time of 220 minutes, this stage has not yet been reached, which allows tuning of the NW aspect ratio, which can be useful for device fabrication. As discussed before, the aspect ratio of the NWs at an early growth stage is mainly determined by the HS. As can be seen in Figure 3 (b,d), at 60 minutes, the average aspect ratio of the faceted objects is larger for the small holes than for the large holes, even though at t=15 minutes, multiple small faceted objects with a larger aspect ratio are present within the larger holes. This indicates that the merging and subsequent recrystallization leads to a decrease in the aspect ratio. When the faceted objects are merging, the expected shape is a cube as all facets are of the (100) family. For the small hole sizes, there is only a single particle within the hole. Therefore, this re-crystallization effect does not occur and the nanowires can have a higher aspect ratio. The unusually large aspect ratio at the early growth stage of the nanowires that do not merge is not yet fully understood and is subject to further study.

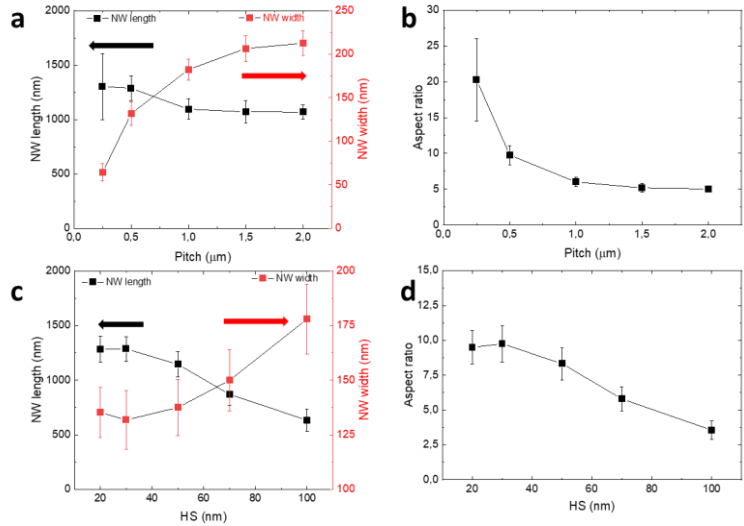

Fig 4: Average PbSnTe nanowire morphology for varying pitch. (a) The average nanowire length (black) and nanowire width (red) and (b) the average aspect ratio for varying pitch size with a constant HS of 30 nm. Lengths and widths were determined from a fitting procedure on fields of nanowires. (c) The average nanowire length (black) and nanowire width (red) and (d) the average nanowire aspect ratio for varying HS at a constant pitch of 500 nm.

Finally, we have studied the elemental composition and distribution within the $Pb_{1-x}Sn_xTe$ NWs. EDX mappings have been acquired on multiple NWs and on a cross-sectional TEM sample fabricated using a focused ion beam (FIB). Based on the individual fluxes of the incoming material, a composition of x=0.5 is expected. However, the elemental mapping of the NW shows a gradient in the composition from x=0.6 to x=0.45 from the bottom to the top (Fig 5 (a,b)). This behavior is consistent for all NWs studied in side-view along the [010] zone axis. Furthermore, the elemental mapping of the NW cross-section also shows a composition gradient with a Sn-rich core (Fig 5 (c,d). Regarding the origin of this phenomenon, we believe that these gradients are caused by solid-state diffusion during growth, where Pb atoms are replaced by Sn atoms, resulting in a Sn-rich core. However, a more extensive study must be carried out to substantiate this hypothesis.

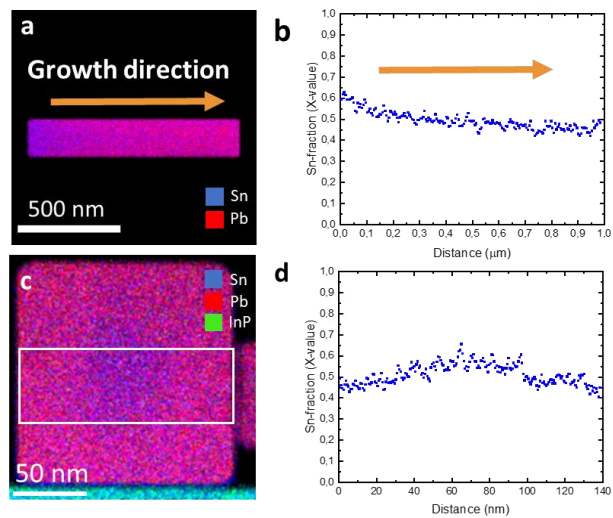

Fig 5: (a) EDX elemental mapping of Pb and Sn of a PbSnTe nanowire. (b)The average Sn-fraction *x* of the nanowire along the growth direction of the NW, quantified and integrated over the entire NW. (c) EDX elemental mapping of Sn, Pb and InP of Figure 1 (e). (d) The Sn-fraction *x* across the width of the cross-sectional area of the NW, quantified and integrated over the white box in Figure 5 (b).

In summary, we have demonstrated the controlled growth of catalyst-free PbSnTe nanowires by a VS mechanism using an optimized pre-deposition procedure. Importantly, the NW aspect ratio can be tuned by the hole size and the pitch. The ability to precisely tune the diameter and the bulk/surface ratio of the NWs is important for elucidating the character of predicted topological surface states in future transport studies.

**Acknowledgments**

M.M. and X.G. contributed equally to this work. The authors thank NanoLab@TU/e for the use of their facilities and their help and support. Additional thanks go to Wim Arnold Bik for helping provide a standard-less calibration for the EDX analysis and Jason Jung (TU/e) and Hans Bolten (Eurofins Materials Science) for their help in sample preparation for cross-sectional TEM. Peter Graat (Eurofins Materials Science) is acknowledged for the in-depth discussions on EDX quantification. This work was supported by the European Research Council (ERC TOCINA 834290) and the Dutch government (OCENW.GROOT.2019.004). The authors recognize Solliance and the Dutch province of Noord Brabant, for funding the TEM facility.

**Data Availability Statement**

The data that supports the findings of this study are openly available in Zenodo at https://doi.org/10.5281/zenodo.10263058.

**Author contributions**

M.M. and X.G. carried out the MBE growth experiments and subsequent SEM analysis. P.L wrote the image analysis software used to extract the NW morphology from SEM images. M.V. performed TEM analysis. E.B. and X.G. supervised the project. M.M., X.G., M.V. and E.B. contributed to the writing of this manuscript,

**Competing interests**

The authors declare no competing interests.